\documentclass[prl,showpacs,reprint,superscriptaddress]{revtex4-1}
\usepackage[utf8]{inputenc}
\usepackage{hyperref,amsmath,amssymb,amsbsy,bm,bbold,color,graphics,graphicx}

\begin{document}
\title{Evidence for a Magnetic Seebeck effect}
\author{Sylvain D. Brechet}
\email{sylvain.brechet@epfl.ch}
\affiliation{Institute of Condensed Matter Physics, Station 3, Ecole Polytechnique Fédérale de Lausanne - EPFL, CH-1015 Lausanne, Switzerland}
\author{Francesco A. Vetro}
\affiliation{Institute of Condensed Matter Physics, Station 3, Ecole Polytechnique Fédérale de Lausanne - EPFL, CH-1015 Lausanne, Switzerland}
\author{Elisa Papa}
\affiliation{Institute of Condensed Matter Physics, Station 3, Ecole Polytechnique Fédérale de Lausanne - EPFL, CH-1015 Lausanne, Switzerland}
\author{Stewart E. Barnes}
\affiliation{James L. Knight Physics Building, 1320 Campo Sano Ave., University of Miami, Coral Gables, FL 33124, USA}
\author{Jean-Philippe Ansermet}
\affiliation{Institute of Condensed Matter Physics, Station 3, Ecole Polytechnique Fédérale de Lausanne - EPFL, CH-1015 Lausanne, Switzerland}
\begin{abstract}
The irreversible thermodynamics	of a continuous medium with magnetic dipoles predicts that a temperature gradient in the presence of magnetisation waves induces a magnetic induction field, which is the magnetic analog of the Seebeck effect. This thermal gradient modulates the precession and relaxation. The Magnetic Seebeck effect implies that magnetisation waves propagating in the direction of the temperature gradient and the external magnetic induction field are less attenuated, while magnetisation waves propagating in the opposite direction are more attenuated.
\end{abstract}
\maketitle
%


The discovery of the spin Seebeck effects in ferromagnetic metals~\cite{Uchida:2008}, in semiconductors~\cite{Jaworski:2010}, and in insulators~\cite{Uchida:2010}, has generated much research for spin transport in ferromagnetic samples of macroscopic dimensions subjected to temperature gradients. The interplay of spin, charge and heat transport defines the rich field known as spin caloritronics~\cite{Bauer:2012}. Prompted by these recent developments, we established a formalism describing the irreversible thermodynamics of a continuous medium with magnetisation~\cite{Brechet:2013}. 

In this letter, we test a particular experimental prediction of this formalism on a YIG slab. We argue that the thermodynamics of irreversible processes implies the existence of a magnetic counter-part to the well-known Seebeck effect. We show how a thermally induced magnetic field modifies the Landau-Lifshitz equation and provide experimental evidence for the Magnetic Seebeck effect by the propagation of magnetisation waves in thin crystals of YIG. The effect of a temperature gradient on the dynamics of the magnetisation on a YIG slab with and without Pt stripes was investigated recently by Obry et al.~\cite{Hillebrands:2012}, Cunha et al.~\cite{Cunha:2013}, Silva et al.~\cite{DaSilva:2013}, Padron-Hernandez et al.~\cite{Padron:2012,Padron:2011b}, Jungfleisch et al.~\cite{Jungfleisch:2013} and Lu et al.~\cite{Lu:2012}.

In general, irreversible thermodynamics predicts couplings between current and force densities. In equation (86) of reference~\cite{Brechet:2013}, we identified the magnetisation force term $\mathbf{m}\,\bm{\nabla}\,\mathbf{B}$. For an insulator like YIG, there is no charge current. As explained in detail in reference~\cite{Brechet:2013}, the transport equation (94) of~\cite{Brechet:2013} implies that the magnetisation force density $\mathbf{M}\,\bm{\nabla}\,\mathbf{B}_{\,\text{ind}}$ induced by a thermal force density $-\,n\,k_B\,\bm{\nabla}\,T$ is proportional and opposite to this force density, i.e.
\begin{equation}\label{lin phen}
\mathbf{M}\,\bm{\nabla}\,\mathbf{B}_{\,\text{ind}} = \lambda\,n\,k_B\,\bm{\nabla}\,T\,,
\end{equation}
which corresponds to equation (155) of reference~\cite{Brechet:2013}, where $\lambda>0$ is a phenomenological dimensionless parameter, $k_B$ is Boltzmann's constant and $n = 1.1\cdot10^{28}\,m^{-3}$ is the Bohr magneton number density of YIG. The thermodynamic formalism does not allow for a direct estimation of $\lambda$. The numerical value of this parameter needs to be evaluated directly from the experimental data, as shown below. 

In the bulk of the sample, as shown in reference~\cite{Brechet:2013}, the magnetisation force density has the structure of a Lorentz force density~\cite{Reuse:2012} expressed in terms of the magnetic bound current density $\mathbf{j}_{\mathbf{M}} = \bm{\nabla}\times\mathbf{M}$~\cite{Griffiths:1999},
\begin{equation}\label{bulk}
\mathbf{M}\,\bm{\nabla}\,\mathbf{B}_{\,\text{ind}} = \mathbf{j}_{\mathbf{M}}\times\mathbf{B}_{\,\text{ind}}\,.
\end{equation}
Thus, using vectorial identities, the phenomenological relations~\eqref{lin phen} and~\eqref{bulk} imply that in the bulk of the system the magnetic induction field $\mathbf{B}_{\,\text{ind}}$, induced by a uniform temperature gradient $\bm{\nabla}\,T$ in the presence of a magnetic bound current density $\bm{\nabla}\times\mathbf{M}$, is given by, i.e.
\begin{equation}\label{B ind}
\mathbf{B}_{\,\text{ind}} = \bm{\varepsilon}_{\mathbf{M}}\times\bm{\nabla}\,T\,,
\end{equation}
where the phenomenological vector $\bm{\varepsilon}_{\mathbf{M}}$ is given by,
\begin{equation}\label{epsilon}
\bm{\varepsilon}_{\mathbf{M}} = -\,\lambda\,n\,k_B\left(\bm{\nabla}\times\mathbf{M}\right)^{-1}\,.
\end{equation}
By analogy with the Seebeck effect, we shall refer to this phenomenon as the Magnetic Seebeck effect. 

The time evolution of the magnetisation $\mathbf{M}$ is given by the Landau-Lifschitz-Gilbert equation, i.e.
\begin{equation}\label{time evolution eq}
\bm{\dot{\mathbf{M}}} = \gamma\,\mathbf{M}\times\mathbf{B}_{\,\text{eff}}-\,\displaystyle{\frac{\alpha}{M_S}}\,\mathbf{M}\times\bm{\dot{\mathbf{M}}}\,,
\end{equation}
where $\gamma$ is the gyromagnetic ratio, $\alpha \simeq 10^{-4}$ is the Gilbert damping parameter of YIG~\cite{Kurebayashi:2011}, $M_S = 1.4\cdot10^{5}\,A\,m^{-1}$ is the magnitude of the effective saturation magnetisation of YIG at room temperature~\cite{Boukchiche:2010}. The effective magnetic induction field $\mathbf{B}_{\,\text{eff}}$ includes the external field $\mathbf{B}_{\,\text{ext}}$, the demagnetising field $\mathbf{B}_{\,\text{dem}}$, the anisotropy field $\mathbf{B}_{\,\text{ani}}$, which behaves as an effective saturation magnetisation in the linear response~\cite{Duncan:1980} and finally a thermally induced field $\mathbf{B}_{\,\text{ind}}$ given by the relation~\eqref{B ind}. The exchange field $\mathbf{B}_{\,\text{int}}$~\cite{Kittel:1949} is negligible in the following, as we consider magnetostatic modes~\cite{Serga:2010}. The demagnetising field $\mathbf{B}_{\,\text{dem}}$ breaks the spatial symmetry and generates an elliptic precession cone. After performing the linear response of the magnetisation in the presence of a thermally induced field $\mathbf{B}_{\,\text{ind}}$, we shall describe how the demagnetising field $\mathbf{B}_{\,\text{dem}}$ affects the magnetic susceptibility.

\begin{figure}[!ht]
\includegraphics[scale=0.32]{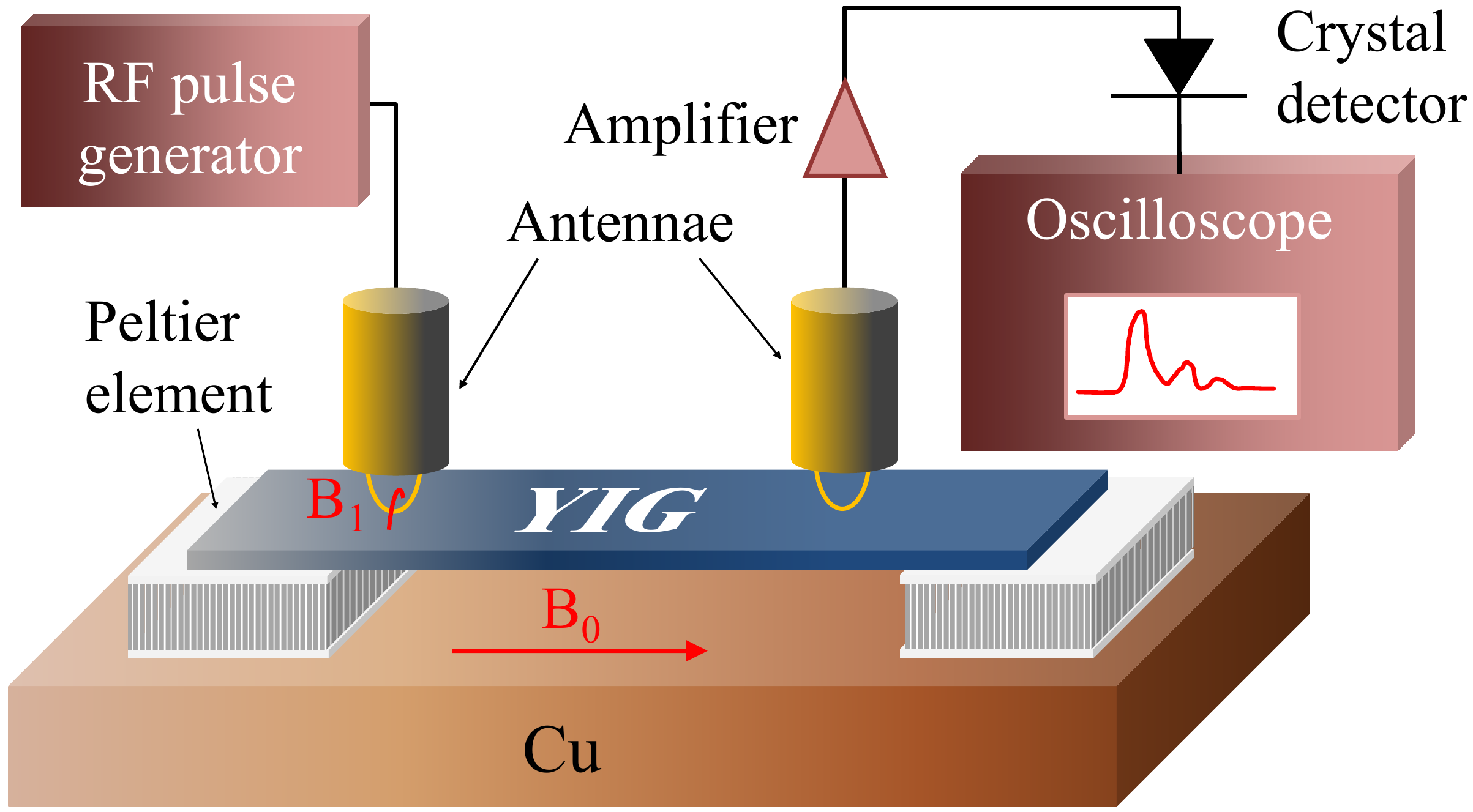}
\caption{Time-resolved transmission measurement of magnetisation waves}
\label{Setup}
\end{figure}

We found evidence for the Magnetic Seebeck effect by exciting locally, at angular frequency $\omega \simeq 2.74\cdot10^{10}\,s^{-1}$, the ferromagnetic resonance of a YIG slab of length $L_z = 10^{-2}\,m$, width $L_y = 2\cdot10^{-3}\,m$ and thickness $L_x = 2.5\cdot10^{-5}\,m$, subjected to a temperature gradient as small as $|\bm{\nabla}\,T| \simeq 2\cdot 10^{3}\,K\,m^{-1}$ generated by Peltier elements. The excitation field is applied on the slab using a local antenna as detailed in reference~\cite{Papa:2013}. For signal transmission experiments, two antennae are used, set approximatively $8\,\text{mm}$ apart, as shown on Fig.~\ref{Setup}. Note that a similar setup for a gradient orthogonal to the YIG slab was investigated recently~\cite{Cunha:2013}. For reasons explained below, these two setups can be expected to probe different mechanisms.

The external magnetic induction field $\mathbf{B}_{\,\text{ext}}$ applied on the YIG film consists of a uniform and constant field $\mathbf{B}_0$ and a small excitation field $\mathbf{b} = b_x\,\mathbf{\hat{x}} + b_y\,\mathbf{\hat{y}}$ locally oscillating in a plane orthogonal to $\mathbf{B}_0 = B_0\,\mathbf{\hat{z}}$. In the limit of a small excitation field, i.e. in the linear limit, the magnetisation field $\mathbf{M}$ consists of a uniform and constant field $\mathbf{M}_S = M_S\,\mathbf{\hat{z}}$ and a response field $\mathbf{m} = m_x\,\mathbf{\hat{x}} + m_y\,\mathbf{\hat{y}}$ locally oscillating in a plane orthogonal to $\mathbf{M}_S$ such that $\mathbf{m}\ll\mathbf{M}_S\,$. The linear response of the magnetisation to the excitation field, according to the time evolution equation~\eqref{time evolution eq} is given by,
\begin{equation}\label{time evolution eq bis}
\bm{\dot{\mathbf{m}}} = \gamma\left(\mathbf{m}\times\mathbf{B}_0 + \mathbf{M}_S\times\mathbf{B}_1\right) -\,\displaystyle{\frac{\alpha}{M_S}}\,\mathbf{M}_S\times\bm{\dot{\mathbf{m}}}\,,
\end{equation}
where the first-order magnetic induction field $\mathbf{B}_1$ yields, 
\begin{equation}\label{B ter}
\mathbf{B}_1 = \mathbf{b} -\,\mu_0\left(\mathbf{k}_T\cdot\bm{\nabla}^{-1}\right)\,\mathbf{m}\,,
\end{equation}
$\mu_0$ is the magnetic permeability of vacuum and the thermal wave vector,
\begin{equation}\label{wave vector}
\mathbf{k}_T = \frac{\lambda\,n\,k_B}{\mu_0\,M_S^2}\,\bm{\nabla}\,T\,.
\end{equation}
To obtain the expressions~\eqref{B ter} and \eqref{wave vector}, we used the linear vectorial identity,
\begin{align*}
&\left(\bm{\nabla}\times\mathbf{M}\right)^{-1}\times\bm{\nabla}\,T = \frac{1}{M_S^2}\left(\bm{\nabla}^{-1}\times\mathbf{m}\right)\times\bm{\nabla}\,T \\
&= \frac{1}{M_S^2}\left(\bm{\nabla}\,T\cdot\bm{\nabla}^{-1}\right)\,\mathbf{m} -\,\frac{1}{M_S^2}\left(\bm{\nabla}\,T\right)\,\bm{\nabla}^{-1}\,\mathbf{m}\,,
\end{align*}
where $\bm{\nabla}^{-1}\cdot\bm{\nabla}=1$ and the last term on the RHS vanishes since it averages out on a precession cycle.

The vectorial time evolution equation~\eqref{time evolution eq bis} is written explicitly in Cartesian coordinates as,
\begin{align}\label{time evol syst}
&\dot{m}_x = \left(\omega_0 + \omega_M\,\mathbf{k}_T\cdot\bm{\nabla}^{-1}\right)m_y + \alpha\,\dot{m}_y -\,\omega_M\,\mu_0^{-1}\,b_y\,,\nonumber\\
&\dot{m}_y = -\,\left(\omega_0 + \omega_M\,\mathbf{k}_T\cdot\bm{\nabla}^{-1}\right)m_x-\,\alpha\,\dot{m}_x + \omega_M\,\mu_0^{-1}\,b_x\,,
\end{align}
where the angular frequencies $\omega_0$ and $\omega_M$ are defined respectively as,
\begin{equation}\label{parameters}
\omega_0 = \gamma\,B_0\,,\qquad \omega_M = \gamma\,\mu_0\,M_S\,.
\end{equation}
In a stationary regime, The magnetic excitation field $\mathbf{b}$ and the magnetisation response $\mathbf{m}$ are oscillating at an angular frequency $\omega$, which is expressed in Fourier series as,
\begin{equation}\label{m}
\begin{split}
&b_x = \sum_{\mathbf{k}}\,b_{\mathbf{k}}\,e^{i\left(\mathbf{k}\cdot\mathbf{x}-\,\omega t + \frac{\pi}{2}\right)}\,,\quad\!\!\!\quad b_y = \sum_{\mathbf{k}}\,b_{\mathbf{k}}\,e^{i\left(\mathbf{k}\cdot\mathbf{x}-\,\omega t\right)}\,,\\
&m_x = \sum_{\mathbf{k}}\,m_{\mathbf{k}}\,e^{i\left(\mathbf{k}\cdot\mathbf{x}-\,\omega t + \frac{\pi}{2}\right)}\,,\quad\!\!\!\! m_y = \sum_{\mathbf{k}}\,m_{\mathbf{k}}\,e^{i\left(\mathbf{k}\cdot\mathbf{x}-\,\omega t\right)}\,,
\end{split}
\end{equation}
where the eigenstates $b_{\mathbf{k}}$ and $m_{\mathbf{k}}$ are complex-valued and dephased. 

The Cartesian components of the eigenmodes $k_{x,y,z}$ satisfy the boundary conditions of null $\mathbf{m}$ at the surface of the sample,
\begin{equation}\label{quantisation}
k_{x,y,z} = \frac{n_{x,y,z}\,\pi}{L_{x,y,z}}\,,
\end{equation}
where $n_{x,y,z}\in\mathbb{N}$~\cite{Papa:2013}.

The eigenstates of the excitation field $b_{\mathbf{k}}$ and the response field $m_{\mathbf{k}}$ are related through the magnetic susceptibility $\chi_{\mathbf{k}}$, i.e.
\begin{equation}\label{mag rel}
m_{\mathbf{k}} = \mu_0^{-1}\,\chi_{\mathbf{k}}\,b_{\mathbf{k}}\,.
\end{equation}
The time evolution equations~\eqref{time evol syst}, the definition~\eqref{parameters}, and the Fourier series~\eqref{m} in the stationary regime imply that the magnetic susceptibility $\chi_{\mathbf{k}}$ is given by,
\begin{equation}\label{susceptibility}
\chi_{\mathbf{k}} = -\,\frac{1}{\Omega -\,\Omega_{0} + i\left(\alpha\,\Omega + \mathbf{k}_T\cdot\mathbf{k}^{-1}\right)}\,,
\end{equation}
where the dimensionless parameter $\Omega$ and $\Omega_{0}$ are respectively defined as,
\begin{equation}\label{wavenumber}
\Omega = \frac{\omega}{\omega_M}\,, \qquad \Omega_{0} = \frac{\omega_0}{\omega_M}\,.
\end{equation}
The demagnetising field $\mathbf{B}_{\,\text{dem}} = -\,\mu_0\,m_x\,\mathbf{\hat{x}}$ causes the damping and the magnetic susceptibility $\chi_{\mathbf{k}x}$ along the $x$-axis to differ respectively from the damping and the magnetic susceptibility $\chi_{\mathbf{k}y}$ along the $y$-axis. The resonance frequency $\sqrt{\omega_0\left(\omega_0 + \omega_M\right)}$ is given by Kittel's formula~\cite{Kittel:2004} to first-order in $\alpha$ and $\mathbf{k}_{T}$. Thus, the magnetic susceptibilities $\chi_{\mathbf{k}x,y}$ yield, 
\begin{equation}\label{susceptibilities}
\chi_{\mathbf{k}x,y} = -\,\frac{1}{\Omega -\,\sqrt{\Omega_{0}\left(\Omega_{0}+1\right)} + i\,r_{x,y}\left(\alpha\,\Omega + \mathbf{k}_T\cdot\mathbf{k}^{-1}\right)}\,,
\end{equation}
where $r_{x,y}>0$ are phenomenological damping scale factors accounting for symmetry breaking. 

As shown by Cunha et al. on Fig.1(a) of reference~\cite{Cunha:2013}, the propagating modes of the magnetisation waves in the bulk of YIG are magnetostatic backward volume modes (MSBVM) propagating in the direction $-\,\mathbf{k}^{-1}$. The expressions~\eqref{susceptibilities} and~\eqref{wave vector} for the magnetic susceptibilities and the thermal wave vector $\mathbf{k}_T$, imply that the magnetisation waves propagating from the cold to the hot side, i.e. $\mathbf{k}_T\cdot\mathbf{k}^{-1}<0$, are less attenuated by the temperature gradient and the magnetisation waves propagating from the hot to the cold side, i.e. $\mathbf{k}_T\cdot\mathbf{k}^{-1}>0$ are further attenuated.

\begin{figure}[!ht]
\includegraphics[scale=0.50]{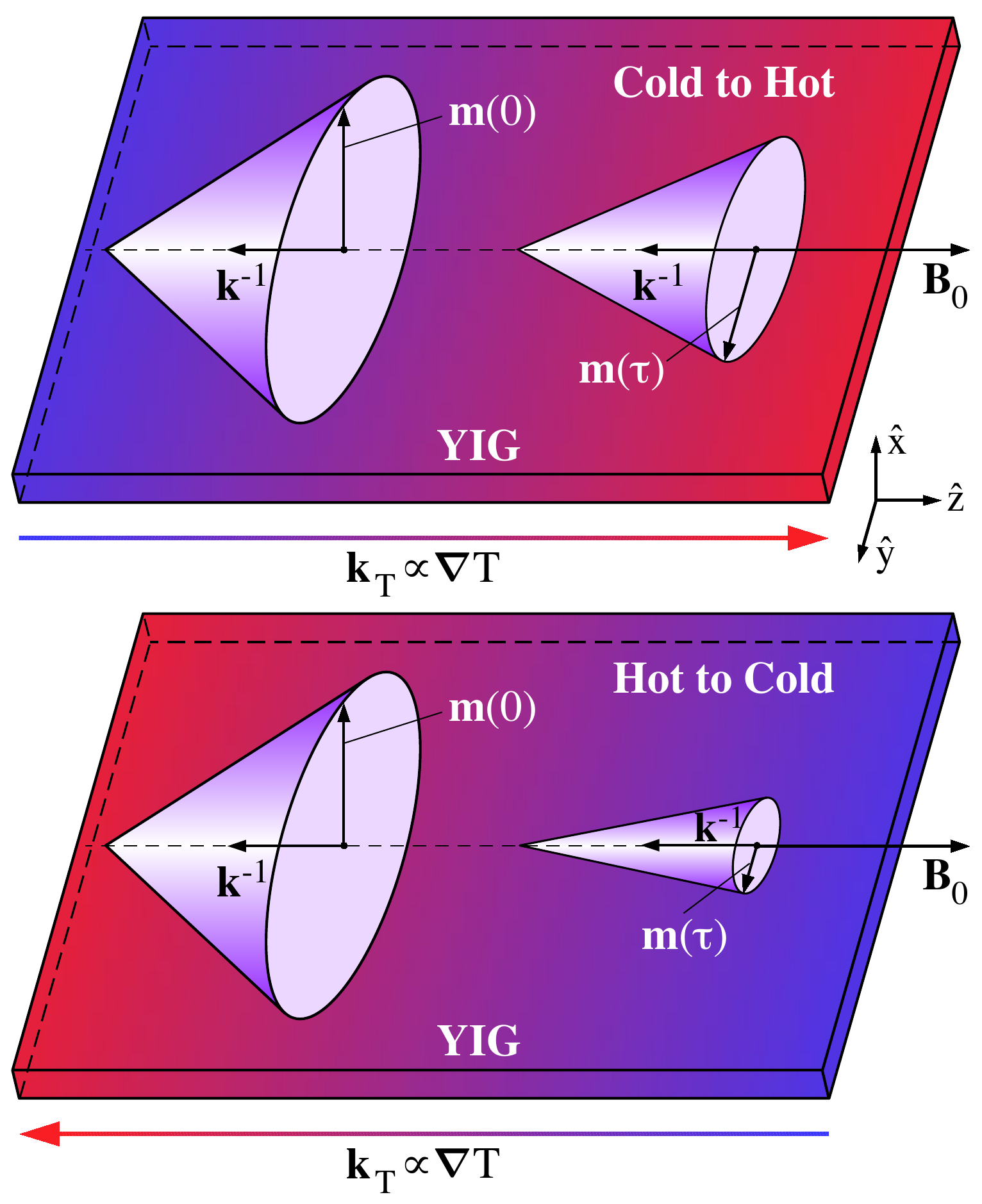}
\caption{Propagation of magnetisation waves from the cold to the hot side (top) and vis versa (bottom). The cones describe the precession of the magnetisation at excitation $\mathbf{m}(0)$ and at detection $\mathbf{m}(\tau)$. The amount of damping depends on the relative orientation $\mathbf{k}_T$ of the temperature gradient with respect to the magnetisation wave propagation direction $-\,\mathbf{k}^{-1}$.}
\label{cone}
\end{figure}

Thus, the opening angle of the precession cone of the magnetisation $\mathbf{m}$ for a magnetisation wave propagating in the direction of the temperature gradient decreases less than the opening angle for a magnetisation wave propagating in the opposite direction, as shown on Figure~\ref{cone}.

\begin{figure}[!ht]
\includegraphics[scale=0.47]{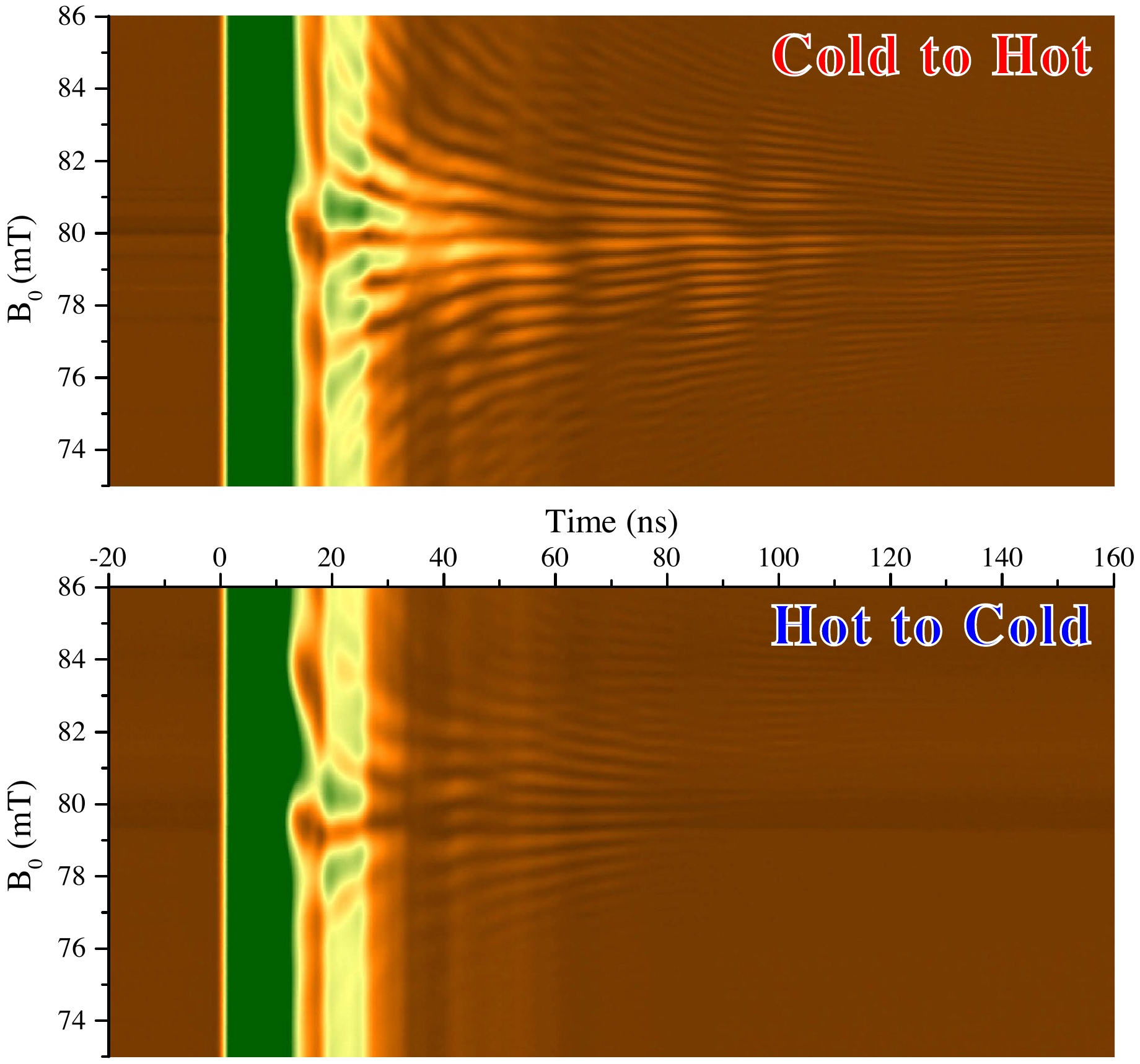}
\caption{Transmitted signals from the cold to the hot side and from the hot to the cold side as a function of the magnetic field $\mathbf{B}_0$ and of the detection time after a $15\,\text{ns}$ pulsed excitation at $4.36\,\text{GHz}$. The lighter areas correspond to a larger signal.}
\label{Propagation_2D}
\end{figure}

This is confirmed experimentally by detecting inductively at one end of the sample the signal that results from an excitation pulse of $15\,\text{ns}$ duration at the other end. The signals obtained by sweeping the magnetic induction field $\mathbf{B}_0$ for the propagation of magnetisation waves from the cold end to the hot end or from the hot end to the cold end are given on Fig.~\ref{Propagation_2D}. Clearly, the waves propagating from the cold to the hot side appear to decay less rapidly than the waves propagating from the hot to the cold side.

\begin{figure}[!ht]
\includegraphics[scale=0.38]{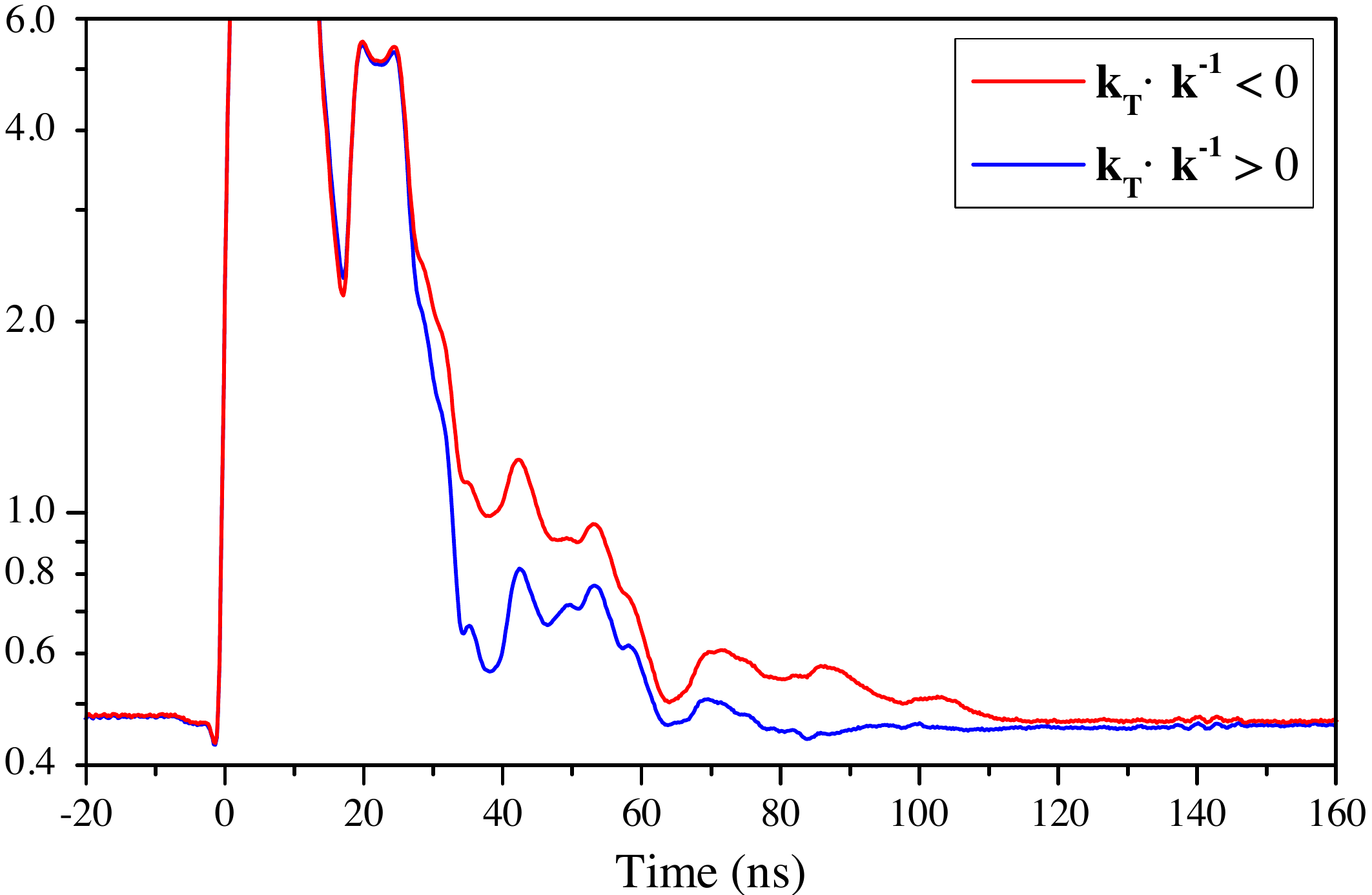}
\caption{Transmitted signal as a function of time after a $15\,\text{ns}$ pulsed excitation at $4.36\,\text{GHz}$.}
\label{Propagation_1D}
\end{figure}

The time evolution of the signals for the waves propagating in the direction of the gradient or opposite to it are obtained by averaging the signals over the range of the magnetic induction field $\mathbf{B}_0$ and displayed on Fig.~\ref{Propagation_1D}. 
The signal is a convolution of $k_z$ modes that have different group velocities and decay exponentially due to the damping. The peaks were identified in reference~\cite{Padron:2011} as the result of the propagation of odd modes. Since the peaks of the transmitted signals are detected at the same time, the temperature gradient does not affect significantly the $k_z$ mode group velocities. Moreover, from the logarithmic scale for the signal on Fig.~\ref{Propagation_1D}, a larger difference in attenuation between the signals for small $k_z$ modes is inferred. This is in line with the theoretical prediction, made by equation~\eqref{susceptibilities}, for the Magnetic Seebeck effect to be proportional to $k_z^{-1}$. Moreover, since the relative difference between the signals is due to the temperature gradient, we can estimate the relative difference between the damping terms $\alpha\,\Omega$ and $\mathbf{k}_T\cdot\mathbf{k}^{-1}$ appearing in the expression~\eqref{susceptibilities} for the magnetic susceptibilities. Comparing the signals at $t=40\,\text{ns}$, we find that the dimensionless parameter $\lambda \simeq 6\cdot 10^{-7}$, which corresponds to a thermal damping ratio $|\mathbf{k}_T\cdot\mathbf{k}^{-1}|/\alpha\,\Omega \simeq 0.3$ less that an order of magnitude below the self-oscillation threshold.

\begin{figure}[!ht]
\includegraphics[scale=0.38]{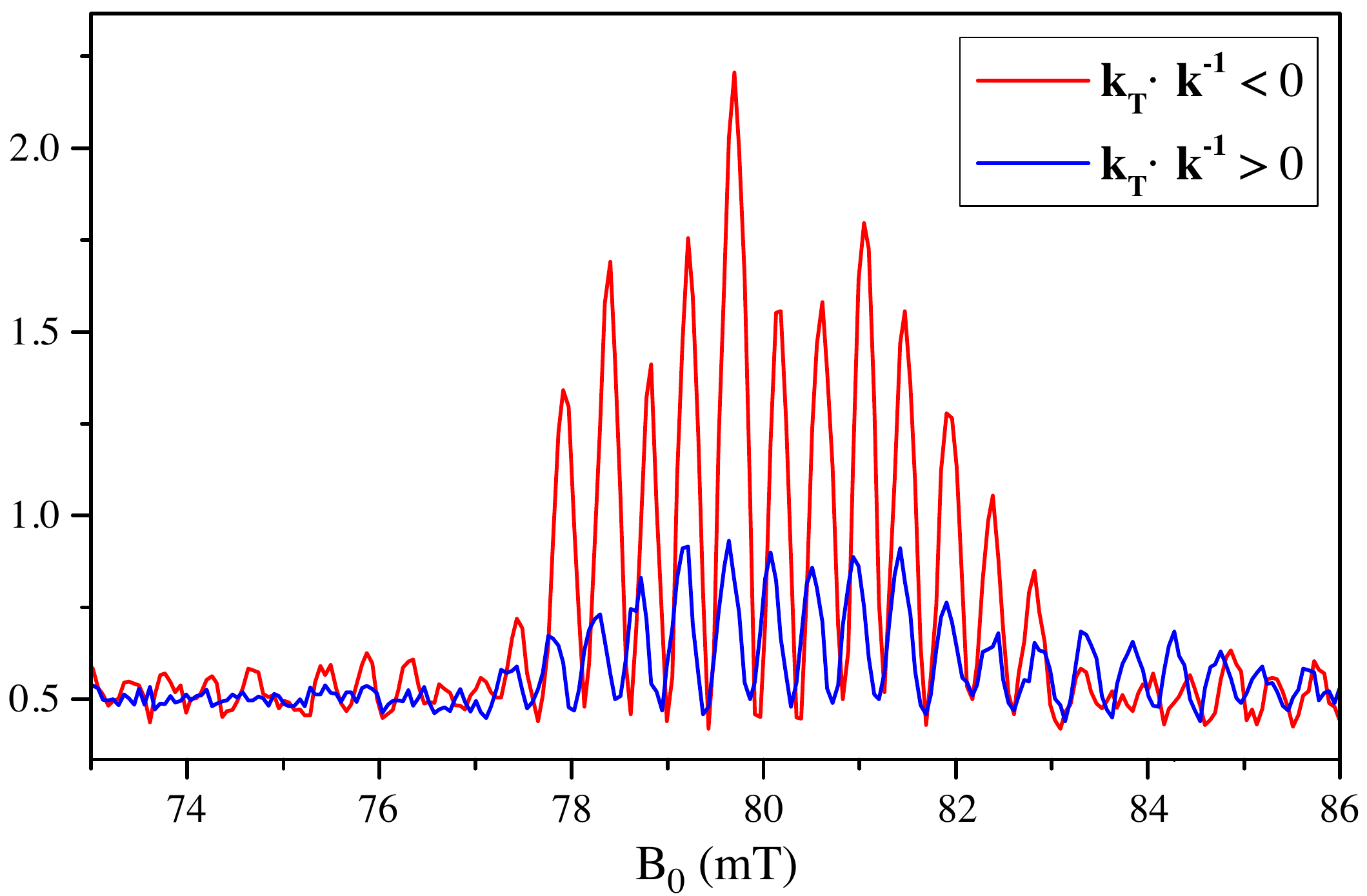}
\caption{FMR signal of a $15\,\text{ns}$ pulsed excitation at $4.36\,\text{GHz}$ detected after $70\,\text{ns}$, after baseline correction.}
\label{FMR}
\end{figure}

The difference in attenuation between the signals is also shown on the FMR spectrum detected $70\,\text{ns}$ after the pulse and displayed on Fig.~\eqref{FMR}. The spectral linewidth $\sim 0.2\,\text{mT}$ corresponds to inhomogeneous broadening, since it is much larger than the homogeneous linewidth $\sim\alpha\,B_{\,\text{eff}}$~\cite{Vonsovskii:1966}.

As rightly pointed out in reference~\cite{Hillebrands:2012}, the temperature dependence of the saturation magnetisation affects the amplitude of the magnetisation waves. However, since our experimental setup is sufficiently close to the self-oscillation threshold for a temperature gradient that is small enough, we expect the dynamic contribution $\mathbf{k}_{T}\cdot\mathbf{k}^{-1}$ to be larger than the static contribution due to the temperature dependence of the saturation magnetisation. Moreover, in contrast to the claim made in reference~\cite{Hillebrands:2012}, Fig.~\ref{Propagation_1D} shows that magnetisation waves can propagate with and against the temperature gradient and that the effect of the temperature is proportional to $k_z^{-1}$.

For a temperature gradient orthogonal to the YIG plane, Cunha et al.~\cite{Cunha:2013} showed that the temperature gradient affects the propagation of magnetisation waves only when Pt is deposited on the YIG slab. The effect is accounted for by a model of spin injection and spin pumping, detailed by Ando et al.~\cite{Ando:2008}, at the interface between Pt and YIG. The quantitative analysis of the data is presented in reference~\cite{DaSilva:2013}. In reference~\cite{Cunha:2013}, it is stated clearly that the effect does not occur in the absence of Pt on the surface. When Pt is removed in such a setup where $\mathbf{k}_T\cdot\mathbf{k}^{-1} = 0$, the mechanism invoked by Cunha et al. is not operative and our mechanism is not effective either.

In summary, we point out that thermodynamics of irreversible processes implies a coupling between heat current and magnetisation precession in a temperature gradient. This effect can be expressed by an induced magnetic field $\mathbf{B}_{\,\text{ind}}$ proportional to the applied temperature gradient. Thus, we suggest to refer to it as a Magnetic Seebeck effect, since it is the magnetic analog of the regular Seebeck effect. It is distinct from the magneto-Seebeck effect, which refers a change in the Seebeck coefficient due to the magnetic response of nanostructures~\cite{Walter:2011}. We analyse how the Landau-Lifshitz equation is modified, and find a contribution to the dissipation that is linear in $\bm{\nabla}\,T$. Hence, this effect can increase or decrease the damping, depending on the orientation of the wave vector of the excited magnetostatic mode with respect to the temperature gradient. If the temperature gradient could be made strong enough, i.e. $\mathbf{k}_T\cdot\mathbf{k}^{-1} > \alpha\,\Omega$, then the damping would be negative and the magnetisation would undergo self-oscillation. This would be analogous to the magnetisation self-oscillation described in chapter $7$ of reference~\cite{Barnes:2012} and the heat-equivalent of Berger's SWASER predicted for charge-driven spin polarised currents~\cite{Berger:1998}.

\begin{acknowledgements}
We thank Fran\c{c}ois A. Reuse, Klaus Maschke and Joseph Heremans for insightful comments and acknowledge the following funding agencies : Polish-Swiss Research Program NANOSPIN PSRP-$045/2010$; Deutsche Forschungsgemeinschaft SS$1538$ SPINCAT, no. AN$762/1$.
\end{acknowledgements}


\bibliography{references}

\end{document}